# Internal gravity waves near to the sources of disturbances at the critical modes of generation

**Vitaly V. Bulatov, Yuriy V. Vladimirov**
**Institute for Problems in Mechanics**
**Russian Academy of Sciences**
**Pr.Vernadskogo 101-1, 117526 Moscow, Russia**
**bulatov@index-xx.ru**

*Abstract*
*The paper presents the description of the structure of the nearest field of the internal gravity waves at the critical modes of their excitation. Studied are the exact solutions both for the elevation component and the vertical component of the speed describing the structure of the wave field in the direct vicinity of the source. At that the single mode of the elevation is expressed through the full elliptic integral of the first order, and the single mode of the vertical speed - through McDonald function and the logarithmic functions. As the result of the study it was possible to obtain expressions for the full field representing the sum of the wave modes and expressed through the derivatives of the gamma function. The obtained asymptotic and exact representations of the solution allow to describe the critical modes of generation of the internal gravity waves near to the sources of excitations - for the wide ranges of the sources movement velocity .*

In the present paper we consider the near field of the internal gravity waves at the critical modes of generation. As we can see from the results gained in the [1-7], in some cases the near wave field has qualitatively different structure and constructing of the exact and asymptotic presentations of the near fields with consideration of the effects of the critical modes of the generation enables to study the physically interesting cases of the wave dynamics of the stratified media.

For constructing the asymptotic presentations of the solutions describing the singularities of the critical modes of generation of the internal waves near to the disturbing sources, similarly to the [1-7] we consider the rectilinear uniform motion of the point-type source of disturbances with the velocity $V$ on the stationary depth: $z_0 = const, y_0 = 0, x_0 = -V\tau$, and the separately taken mode. Using the results from [1-7], we shall proceed from the following expression of the separate mode of elevation $\eta_n$

$$\eta_n(\lambda, y) = \frac{1}{2\pi V} \int_0^\infty \frac{\omega_n^2(k)}{k} \varphi_n(z,k) \frac{\partial \varphi_n(z_0,k)}{\partial z_0} \int_{-\infty}^\lambda \cos(\omega_n(k)(\lambda - \xi)/V) J_0(k\sqrt{y^2 + \xi^2}) d\xi dk \qquad (1)$$

where $x + Vt = \lambda$, $J_0$ - Bessel function of the zeroth order and is considered the stable mode of generation of the internal waves caused by the moving source of the disturbances.

Further we shall consider the behavior of the separate mode of then elevation $\eta_n(\lambda, y)$ at the small values of $\lambda, y$, that is in the vicinity of the moving source of disturbances

$$\eta_n(\lambda, y) = \eta_n(0,0) + T_n \lambda + B_n y + ... \qquad (2)$$
$$T_n = \partial \eta_n(0,0)/\partial \lambda \quad, \quad B_n = \partial \eta_n(0,0)/\partial y$$

It is obvious, that due to the symmetry of the problem with respect to the variable $y$, the function $\eta_n(\lambda, y)$ is even for the given variable, and accordingly, $B_n = 0$

Let's consider further the behavior of function $\eta_n(0,0)$. The inner integral in (1) is taken

$$R_n \equiv \int_0^\infty \cos(\frac{\omega_n(k)\xi}{V}) J_0(k\xi) d\xi = \frac{1}{\mu_n(k)}, k > \frac{\omega_n(k)}{V}$$

$$R_n = 0, k < \frac{\omega_n(k)}{V}$$

$$\mu_n(k) = \sqrt{k^2 - \frac{\omega_n^2(k)}{V^2}}$$

Then

$$\eta_n(0,0) = \int_{d_n}^\infty \frac{\omega_n^2(k)}{k \mu_n(k)} F_n(k, z, z_0) dk$$

$$F_n(k, z, z_0) = \frac{1}{2\pi V} \varphi_n(z, k) \frac{\partial \varphi_n(z_0, k)}{\partial z_0}$$

where $d_n$ - the root of the equation $k^2 V^2 = \omega_n^2(k)$, at $V < c_n$, $c_n = \frac{d\omega_n(k)}{dk}$ (( $k = 0$ ) - the maximal group velocity of the n-th mode $d_n = 0$, at $V > c_n$.

Further, it is analogous to [1-7], for ease and obviousness of the calculations, we shall consider the case of the exponentially stratified liquid $N(z) = const$, then

$$F_n(k, z, z_0) = \frac{1}{V N^2 H^2} \sin(\frac{\pi n z}{H}) \cos(\frac{\pi n z_0}{H}) \equiv A_n$$

Let's consider separately two cases.
The case $V > c_n$. The expression for $\eta_n(0,0)$ will look like

$$\eta_n(0,0) = A_n \int_0^\infty \frac{dk}{\sqrt{k^2 + b_n^2}\sqrt{k^2 + \varepsilon_n^2}} = \frac{A_n}{b_n} K(\frac{\sqrt{b_n^2 - \varepsilon_n^2}}{b_n})$$

$$c_n = \frac{NH}{\pi n}, b_n = \frac{\pi n}{H}$$

where $K(x) = \int_0^{\pi/2} \frac{d\tau}{\sqrt{1 - x^2 \sin^2 \tau}}$ - complete elliptic integral of the first kind,

$b_n^2(1 - \frac{c_n^2}{V^2}) = \varepsilon_n^2$ - the measure of deflection of $V$ from $c_n$

Finally we shall gain

$$\eta_n(0,0) = \frac{1}{\pi N^2 HV} K(\frac{c_n}{V}) \sin(\frac{\pi n z}{H}) \cos(\frac{\pi n z_0}{H})$$

The case $V \langle c_n$.

$$\eta_n(0,0) = A_n \int\limits_{\varepsilon_n}^{\infty} \frac{dk}{\sqrt{k^2+b_n^2}\sqrt{k^2-\varepsilon_n^2}} = \frac{A_n}{\sqrt{b_n^2+\varepsilon_n^2}} K(\frac{b_n}{\sqrt{b_n^2+\varepsilon_n^2}})$$

where $b_n^2(\frac{c_n^2}{V^2}-1) = \varepsilon_n^2$

Finally

$$\eta_n(0,0) = \frac{1}{\pi N^2 H c_n} K(\frac{V}{c_n}) \sin(\frac{\pi n z}{H}) \cos(\frac{\pi n z_0}{H})$$

Let's transfer now to the calculation of coefficient $T_n$ in (2), which one, apparently, due to its determination with accuracy up to $V$, is the value of the separate mode $W_n$ of the vertical velocity at $\lambda = y = 0$

$$\frac{\partial \eta_n}{\partial \lambda} = \frac{1}{V}\frac{\partial \eta_n}{\partial t} = \frac{1}{V} W_n$$

$$T_n = \frac{W_n(0,0)}{V}$$

$$\frac{\partial \eta_n(0,0)}{\partial \lambda} = \frac{1}{2\pi V} \int\limits_0^{\infty} \frac{\omega_n^2(k)}{k} \varphi_n(z,k) \frac{\partial \varphi_n(z_0,k)}{\partial z_0} J_0(k\sqrt{y^2+\lambda^2}) dk -$$

$$- \frac{1}{2\pi V} \int\limits_0^{\infty} \frac{\omega_n^2(k)}{k} \varphi_n(z,k) \frac{\partial \varphi_n(z_0,k)}{\partial z_0} \int\limits_{-\infty}^{\lambda} \frac{\omega_n(k)}{V} \sin(\omega_n(k)(\lambda-\xi)/V) J_0(k\sqrt{y^2+\xi^2}) d\xi dk =$$

$$= P_{1n} - P_{2n}$$

Let's consider the item $P_{1n}$. For the case $N(z) = const$ we have

$$P_{1n}(\lambda, y) = \frac{1}{2\pi V} \int\limits_0^{\infty} \frac{k N^2}{k^2+b_n^2} \varphi_n(z,k) \frac{\partial \varphi_n(z_0,k)}{\partial z_0} J_0(k\sqrt{y^2+\lambda^2}) dk =$$

$$= \frac{n}{H^2 V} \sin(\frac{\pi n z}{H}) \cos(\frac{\pi n z_0}{H}) K_0(\frac{\pi n}{H}\sqrt{y^2+\lambda^2})$$

where $K_0(x)$ - McDonald's function of the zeroth order. The series $\sum\limits_{n=1}^{\infty} P_{1n}$ is possible to sum

$$P_1(\lambda, y) = \sum\limits_{n=1}^{\infty} P_{1n}(\lambda, y) =$$

$$= \frac{1}{4\pi V} \left\{ \frac{z_-}{(r^2+z_-^2)^{3/2}} + \frac{z_+}{(r^2+z_+^2)^{3/2}} - \sum\limits_{m=1}^{\infty} \left[ \frac{2mH-z_-}{(r^2+(2mH-z_-)^2)^{3/2}} - \right.\right.$$

$$\left.\left. - \frac{2mH+z_-}{(r^2+(2mH+z_-)^2)^{3/2}} + \frac{2mH-z_+}{(r^2+(2mH-z_+)^2)^{3/2}} - \frac{2mH+z_+}{(r^2+(2mH+z_+)^2)^{3/2}} \right] \right\}$$

where $r = \sqrt{\lambda^2+y^2}$, $z_- = z-z_0$, $z_+ = z+z_0$

At $r=0$ the given a series also is summed up and expresses through the derivative $\Psi'(x)$, where $\Psi(x)=(\ln \Gamma(x))'$ - the psi-function, or the derivative of the gamma-function.

$$P_1(0,0) = \frac{1}{16\pi H^2 V}(\Psi'(-\frac{z_-}{2H}) + \Psi'(-\frac{z_+}{2H}) -$$

$$-\Psi'(\frac{z_-}{2H}) - \Psi'(\frac{z_+}{2H}) + \frac{4H^2}{z_-^2} + \frac{4H^2}{z_+^2})$$

Let's transfer to considering the item $P_{2n}(0,0)$ ( $N(z) = const$ )

$$P_{2n}(0,0) = \frac{nN}{H^2 V^2}\sin(\frac{\pi n z}{H})\cos(\frac{\pi n z_0}{H})\int_0^\infty \frac{k^2}{(k^2+b_n^2)^{3/2}}\int_0^\infty \sin(\frac{\omega_n(k)\xi}{V})J_0(k\xi)dkd\xi \quad (3)$$

The inner integral in (3) is taken:

$$Q_n \equiv \int_0^\infty \sin(\frac{\omega_n(k)\xi}{V})J_0(k\xi)d\xi = \frac{1}{\mu_n(k)}, \frac{\omega_n(k)}{V} > k$$

$$Q_n = 0, \frac{\omega_n(k)}{V} < k$$

Here $\mu_n(k) = \sqrt{\frac{\omega_n^2(k)}{V^2} - k^2}$. Further we shall gain

$$P_{2n}(0,0) = \frac{nN}{H^2 V^2}\sin(\frac{\pi n z}{H})\cos(\frac{\pi n z_0}{H})\int_0^{\varepsilon_n} \frac{kdk}{(k^2+b_n^2)\sqrt{\varepsilon_n^2 - k^2}}$$

where $b_n^2(\frac{c_n^2}{V^2} - 1) = \varepsilon_n^2$

Finally we have

$$P_{2n}(0,0) = \frac{nN}{H^2 V^2}\sin(\frac{\pi n z}{H})\cos(\frac{\pi n z_0}{H})\ln(\frac{c_n}{V} + \sqrt{\frac{c_n^2}{V^2} - 1})$$

It is interesting to note, that the series $\sum_{n=1}^\infty P_{2n}(0,0)$, due to the properties of integrals $Q_n$ and decreasing with the number of the mode of the values of the maximal group velocities $c_1 > c_2 > c_3 > ...$ has the final number of nonzero items.

Thus, gained asymptotic and exact presentations of the solution allow to describe the critical modes of generation of the internal gravity waves near to the sources of disturbances for the broad ranges of the velocities of movement of the sources of disturbances.